\documentclass[aps,prd,nofootinbib]{revtex4}
\usepackage{graphicx}

\begin{document}

\title{Can massless wormholes mimic a Schwarzschild black hole in the strong field lensing?}
\author{Ramil N. Izmailov}
\email{izmailov.ramil@gmail.com}
\affiliation{Zel'dovich International Center for Astrophysics, Bashkir State Pedagogical University, 3A, October Revolution Street, Ufa 450008, RB, Russia}
\author{Amrita Bhattacharya}
\email{amrita_852003@yahoo.co.in}
\affiliation{Department of Mathematics, Kidderpore College, 2, Pitamber Sircar Lane, Kolkata 700023, WB, India}
\author{Eduard R. Zhdanov}
\email{zhdanov@ufanet.ru}
\affiliation{Zel'dovich International Center for Astrophysics, Bashkir State Pedagogical University, 3A, October Revolution Street, Ufa 450008, RB, Russia}
\author{Alexander A. Potapov}
\email{potapovaa2008@rambler.ru}
\affiliation{Department of Physics \& Astronomy, Bashkir State University, 49, Lenin Street, Sterlitamak 453103, RB, Russia}
\author{K.K. Nandi}
\email{kamalnandi1952@yahoo.co.in}
\affiliation{Zel'dovich International Center for Astrophysics, Bashkir State Pedagogical University, 3A, October Revolution Street, Ufa 450008, RB, Russia}
\affiliation{High Energy and Cosmic Ray Research Center, University of North Bengal, Siliguri 734 013, WB, India}

\begin{abstract}
Recent trend of research indicates that not only massive but also massless (asymptotic Newtonian mass zero) wormholes can reproduce post-merger initial ring-down gravitational waves characteristic of black hole horizon. In the massless case, it is the non-zero charge of other fields, equivalent to what we call here the "Wheelerian mass", that is responsible for mimicking ring-down quasi-normal modes. In this paper, we enquire whether the same Wheelerian mass can reproduce black hole observables also in an altogether different experiment, viz., the strong field lensing. We examine two classes of massless wormholes, one in the Einstein-Maxwell-Dilaton (EMD) theory and the other in the Einstein-Minimally-coupled-Scalar field (EMS) theory. The observables such as the radius of the shadow, image separation and magnification of the corresponding Wheelerian masses are compared with those of a black hole (idealized SgrA* chosen for illustration) assuming that the three types of lenses share the same minimum impact parameter and distance from the observer. It turns out that, while the massless EMS\ wormholes can closely mimic the black hole in terms of strong field lensing observables, the EMD wormholes show considerable differences due to the presence of dilatonic charge. The conclusion is that masslessless alone is enough to closely mimic Schwarzschild black hole strong lensing observables in the EMS theory but not in the other, where extra parameters also influence those observables. The motion of timelike particles is briefly discussed for completeness.
\end{abstract}

\maketitle


\section{Introduction}
\label{Intro}
Wormhole spacetimes predicted by gravitational theories have non-trivial topological structures and their observational aspects can be one of the interesting topics to understand our Universe. Although wormholes require exotic matter and the formation mechanism of wormholes is not well understood, these objects have not been ruled out by observations to date. In fact, recently some remarkable developments have taken place: It has been shown that the gravity waves from a particle near a phantom thin-shell wormhole surgically created by matching two copies of Schwarzschild black hole can mimic the time domain of quasinormal ringing of the Schwarzschild black hole at early times and differ from it only at sufficiently late times \cite{1,2}. Such a possibility was previously discussed in \cite{3} in connection with an analytic wormhole model. For some recent works relating to this wormhole, see \cite{4,5,6}. Ellis-Bronnikov wormhole of the EMS theory\footnote{It should be mentioned here that Ellis \cite{17} derived two classes of asymptotically flat solutions, one showing naked singularity and the other a regular massive traversable wormhole, which we call here the massive Ellis-Bronnikov wormhole since Bronnikov \cite{18} independently derived it [see the metric (42-45) below]. Interestingly, the two Ellis classes are not really independent but can be connected by a complex Wick rotation, as shown in \cite{19}.} (hereinafter EMS wormhole) have been considered in \cite{7}, who showed that the massless wormhole, depending on the values of its parameters, either rings exclusively as a black hole at all times or rings differently at all times. (By the term massless wormhole, we shall always mean that its asymptotic Newtonian mass of the parent metric is zero). Ringing by \textit{massive} EMS wormhole mimicking Schwarzschild black hole has been analyzed from a different approach in \cite{8}.

An important factor concerning wormholes in general is its stability. While thin-shell wormholes are stable within certain parameter limits, the stability of massless EMS wormhole sourced by ghost field seems to be still an open question: It has been argued that the wormhole is stable under arbitrary space-time perturbations \cite{9}. It has also been shown that the same wormhole metric can also be generated by a special type of non-ghost equation of state and that the wormhole is stable against spherically symmetric and non-spherical axial perturbations \cite{10,11}. However, arguments against the stability of massless EMS wormhole persist showing that the instability leads either to scalar field inflation or to collapse to black hole \cite{12}. Some authors also argue for instability of the massive EMS wormhole collapsing into a Schwarzschild black hole with the exotic scalar field radiated away \cite{13}. Mathematically, it means that the massive wormhole parameter $\gamma$ has to drastically cross over from a real value to a complex value ($\gamma =-i$) during collapse to a final black hole \cite{8}. While the collapse scenario under arbitrary perturbations is not contended, it is unclear what physical phenomenon would mark the cross-over from the real to complex value of $\gamma$ during the evolution of collapse. With all these, it is fair to say that the stability of EMS wormhole, massive or massless, is not yet completely understood for arbitrary perturbations even in the linear approximation, as argued in \cite{7}.

Returning to the main topic of the paper, we note that the gravitational lensing phenomenon is an important tool for understanding the nature of the lens that could be a black hole, wormhole or even a naked singularity. However, there is a fundamental difference between weak and strong field lensing. Strong field light deflection by a Schwarzschild black hole or by a spherically symmetric wormhole displays a logarithmic divergence at the photon sphere \cite{14,15}, where light rays get captured (making infinite number of loops on it) demarcating the strong field limit for those lenses. There is no way that the strong field deflection angle could be expanded to yield the known weak field deflection terms supposed to be valid far away from the photon sphere. This is so because, as shown in \cite{16}, even the zeroth order affine parametric expansion of the exact deflection angle already shows a logarithmic divergence. On the other hand, while the weak field light deflection is useful in itself, strong field gravitational lensing is expected to provide a completely different genre of observables\ for different types of lenses. Also, wormholes mimicking post-merger quasi-normal mode ringing \cite{1,2,3,4,5,7,8} provides a strong motivation for enquiring whether or not massless wormholes of different theories can mimic black hole lensing observables.

Despite having zero asymptotic Newtonian masses, the wormholes are sourced by other types of fields, such as scalar or dilaton having energy, hence an equivalent non-zero mass can be attributed as secondary source that can curve the spacetime causing deflection of light and consequent lensing effects. For instance, the massless EMS wormhole \cite{17,18} is sourced entirely by an exotic (negative energy density) scalar field (a.k.a. ghost field), with the wormhole structure consisting of an exterior positive and an interior negative mass, both of equal magnitude symmetrically lying on either side of the throat. We call each of them a \textit{Wheelerian mass} following "Wheeleresque mantra: Mass without mass \cite{20}" as humorously phrased in \cite{21}. The positive Wheelerian mass causing the light rays to bend toward the mass eventually leads to observable lensing effects. The existence of such non-zero Wheelerian mass then allows us to compare\ how the strong lensing properties of massless wormholes in the two theories, EMD and EMS, differ between themselves as well as with those of the Schwarzschild black hole. Wheelerian mass is a useful concept already used in astrophysical application accounting for dark halo objects in the exterior of our galaxy (see \cite{22}).

The purpose of the present paper is to investigate the strong field lensing observables caused by lightlike motion in the massless EMD and EMS wormholes using Bozza's method \cite{14}. We shall enquire if these observables can mimic those of Schwarzschild black hole assuming that the three types of lenses share the same minimum impact parameter $u_{m}$ and distance $D_{\textmd{\scriptsize{OL}}}$ from the observer. The stability of circular orbits of timelike particles is briefly discussed for completeness.

The paper is organized as follows. A brief preview of the method for obtaining strong field lensing observables would be worthwhile and is presented in Sec.2. The method is applied to massless EMD wormhole in Sec.3 and to EMS wormhole in Sec.4. Numerical estimates are presented in Sec.5 and a brief analysis of the motion of timelike particles is added in Sec.6. Sec.7 concludes the paper. We take units such that $8\pi G=1$, $c=1$.

\section{Bozza's method: A brief preview}
\label{Sec.2}
This method has by now gained considerable attention for its usefulness, so the purpose of this preview is to let the readers readily see what quantities have been calculated to get to the final lensing observables. The method starts with a generic spherically symmetric static spacetime
\begin{equation}
ds^{2}=A(x)dt^{2}-B(x)dx^{2}-C(x)\left( d\theta ^{2}+\sin ^{2}\theta \phi^{2}\right).
\end{equation}%
The equation [34,35]
\begin{equation}
\frac{C^{\prime }(x)}{C(x)}=\frac{A^{\prime }(x)}{A(x)}
\end{equation}%
is assumed to admit at least one positive root and the largest root is called the radius of the photon sphere $x_{m}$. The strong field expansion will take the photon sphere radius as the starting point, which is required to exceed the horizon radius of a black hole or throat radius of a wormhole as the case may be. A light ray coming in from infinity will reach the closest approach distance $x_{0}$ from the centre of the gravitating source before emerging in another direction. By the conservation of angular momentum, $x_{0}$ is related to the impact parameter $u$ by
\begin{equation}
u=\sqrt{\frac{C_{0}}{A_{0}}},
\end{equation}%
where the subscript $0$ indicates that the function is evaluated at $x_{0}$. The minimum impact parameter is defined by%
\begin{equation}
u_{m}=\sqrt{\frac{C_{m}}{A_{m}}},
\end{equation}%
where $C_{m}\equiv C(x_{m})$ etc. From the null geodesics, the deflection angle $\alpha (x_{0})$ can then be calculated as a function of the closest approach [36]:
\begin{equation}
\alpha (x_{0})=I(x_{0})-\pi,
\end{equation}
\begin{equation}
I(x_{0})=\int\limits_{x_{0}}^{\infty }\frac{2\sqrt{B}dx}{\sqrt{C}\sqrt{\frac{C}{C_{0}}\frac{A_{0}}{A}-1}}.
\end{equation}

In the weak field limit of deflection, the integrand is expanded to any order in the gravitational potential and integrated. When we decrease the impact parameter $u$ (and consequently $x_{0}$), the deflection angle increases. Decreasing $u$ further bringing the ray infinitesimally closer to the photon sphere will cause the ray to wind up a large number of times before emerging out. Finally, at $x_{0}=x_{m}$, corresponding to an impact parameter $u=u_{m}$, the deflection angle will diverge and the ray will be captured, i.e., it will wind around the photon sphere indefinitely.

It has been shown that this divergence is logarithmic for all spherically symmetric metrics, which yields an analytical expansion for the deflection angle close to the divergence in the form \cite{14}
\begin{equation}
\alpha (x_{0})=-a\log \left( \frac{x_{0}}{x_{m}}-1\right) +b+O\left(x_{0}-x_{m}\right).
\end{equation}%
The coefficients $a,b$ depend on the metric functions evaluated at $x_{m}$, and the Eq.(7) is defined as the \textit{strong field limit} of the light deflection angle. The coefficients $a,b$ are respectively redefined to $\overline{a}$ and $\overline{b}$ that are obtained as follows. Define two new variables
\begin{eqnarray}
&&y=A(x), \\
&&z=\frac{y-y_{0}}{1-y_{0}},
\end{eqnarray}%
where $y_{0}=A_{0}$. The integral (6) then becomes
\begin{equation}
I(x_{0})=\int\limits_{0}^{1}R(z,x_{0})f(z,x_{0})dz,
\end{equation}
\begin{equation}
R(z,x_{0})=\frac{2\sqrt{By}}{CA^{\prime }}\left( 1-y_{0}\right) \sqrt{C_{0}},
\end{equation}
\begin{equation}
f(z,x_{0})=\frac{1}{\sqrt{y_{0}-\left[ \left( 1-y_{0}\right) z+y_{0}\right]\frac{C_{0}}{C}}},
\end{equation}%
where all functions without the subscript $0$ are evaluated at $x=A^{-1}\left[ \left( 1-y_{0}\right) z+y_{0}\right] $. The function $R(z,x_{0})$ is regular for all values of $z$ and $x_{0}$, while $f(z,x_{0})$ diverges for $z\rightarrow 0$, where
\begin{equation}
f(z,x_{0})\sim f_{0}(z,x_{0})=\frac{1}{\sqrt{\alpha z+\beta z^{2}}},
\end{equation}
\begin{equation}
\alpha =\frac{1-y_{0}}{C_{0}A_{0}^{\prime }}\left( C_{0}^{\prime}y_{0}-C_{0}A_{0}^{\prime }\right),
\end{equation}
\begin{equation}
\beta =\frac{\left( 1-y_{0}\right) ^{2}}{2C_{0}^{2}{A_{0}^{\prime }}^{3}}%
\left[ 2C_{0}C_{0}^{\prime }{A_{0}^{\prime }}^{2}+ \left( C_{0}C_{0}^{\prime \prime }-2{C_{0}^{\prime }}^{2}\right)
y_{0}A_{0}^{\prime }-C_{0}C_{0}^{\prime }y_{0}A_{0}^{\prime \prime }\right] ,
\end{equation}
where primes denote differentiation with respect to $x$.

For the calculation of lensing observables, note that the angular separation of the image from the lens is $\tan \theta = \frac{u}{D_{\textmd{\scriptsize{OL}}}}$, where $D_{\textmd{\scriptsize{OL}}}$ is the distance between the observer and the lens. Specializing to the photon sphere $x_{0}=x_{m}$, the deflection angle in Eq.(7) can be rewritten into a final form
\begin{eqnarray}
\alpha (\theta ) &=&-\overline{a}\log \left( \frac{u}{u_{m}}-1\right) +\overline{b}, \\
u &\simeq &\theta D_{\mathrm{OL}}\textmd{ (assuming small }\theta)
\end{eqnarray}
\begin{equation}
\overline{a}=\frac{a}{2}=\frac{R(0,x_{m})}{2\sqrt{\beta _{m}}}, \; \beta_{m}=\beta |_{x_{0}=x_{m}},
\end{equation}
\begin{eqnarray}
\overline{b} &=&-\pi +b_{R}+\overline{a}\log {\frac{2\beta _{m}}{y_{m}}}, \; y_{m}=A(x_{m}), \\
b_{R} &=&\int_{0}^{1}g(z,x_{m})dz, \\
g(z,x_{m}) &=&R(z,x_{m})f(z,x_{m})-R(0,x_{m})f_{0}(z,x_{m}).
\end{eqnarray}

The impact parameter $u$ is related to the angular separation $\theta $ of images by the relationship given in Eq.(17). Using this, Bozza proposed three strong lensing observables as \cite{14}
\begin{eqnarray}
\theta _{\infty } &=&\frac{u_{m}}{D_{\textmd{\scriptsize{OL}}}}, \\
s &=&\theta _{1}-\theta _{\infty }=\theta _{\infty }\exp \left( \frac{\bar{b}}{\bar{a}}-\frac{2\pi }{\bar{a}}\right) , \\
r &=&2.5\log _{10}\left[ \ \exp \left( \frac{2\pi }{\bar{a}}\right) \ \right],
\end{eqnarray}%
where $\theta _{1}$ is the angular position of the outermost image, $\theta_{\infty }$ is the asymptotic position approached by a set of images in the limit of a large number of loops the rays make around the photon sphere, $s$ is the angular separation between the outermost images resolved as a single image and the set of other asymptotic images, all packed together. $r$ is ratio between the flux of the first image and the flux coming from all the other images.

We shall calculate in what follows the strong field lensing coefficients $\left\{\overline{a},\overline{b}\right\}$ and observables $\left(\theta_{\infty}, s, r\right)$ applying respectively the formulas (18,19) and (22,23,24) to the massless EMD and EMS wormholes with the understanding, once again, that only the asymptotic Newtonian mass of the parent metric is zero, while the Wheelerian mass is not.

\section{Massless wormhole in EMD theory}
\label{Sec.3}
The wormhole solution in the EMD theory together with its massless limit has been derived in the literature \cite{23} and light deflection in it using the Gauss-Bonnet method has been obtained in \cite{24}. The weak field lensing observables have been recently computed and compared with those of EMS wormhole in \cite{25}. The massive parent metric (redefining $r=x$ in it \cite{23}) is:%
\begin{eqnarray}
A(x)&=& 1/B(x)= \frac{(x-r_{1})(x-r_{2})}{(x+d_{0})(x+d_{1})}, \quad C(x)=(x+d_{0})(x+d_{1}), \\
e^{2\phi} &=&\frac{P(x-\Sigma )}{Q(x+\Sigma )},\quad F_{xt}=\frac{P}{(x+\Sigma )^{2}},\quad F_{\theta \varphi }=P\sin \theta,
\end{eqnarray}
where $P$ is the magnetic charge, $Q$ is electric charge, $\Sigma$ is the dilatonic charge and $r_{1},r_{2},d_{0},d_{1}$ are constants. The metric expands in the weak field as
\begin{eqnarray}
A(x) &=&1-\frac{2M}{x}+O(1/x^{2}), \\
M &=&\frac{r_{1}+r_{2}+d_{0}+d_{1}}{2}
\end{eqnarray}%
is the asymptotic Newtonian mass, which is assumed to be zero under the choice of constants $r_{1}=-r_{2},d_{0}=-d_{1}$. By further redefining $q^{2}=2PQ$ and $k^{2}=\Sigma ^{2}+q^{2}$, and using certain other conditions (see [23,24] for details), the massless EMD wormhole metric components read
\begin{eqnarray}
A(x)&=&\left(1+\frac{q^{2}}{x^{2}}\right)^{-1} \\
B(x)&=&\left( 1+\frac{q^{2}}{x^{2}}\right) \left(1+\frac{k^{2}}{x^{2}}\right)^{-1} \\
C(x)&=&x^{2}+q^{2}.
\end{eqnarray}%

In the "standard" radial coordinate $R=\sqrt{x^{2}+q^{2}}$, the throat appears at $R_{\textmd{\scriptsize{th}}}=q$, which translates to $x_{\textmd{\scriptsize{th}}}=0$. The two functions $R(z,x_{0})$ and $f(z,x_{0})$ work out to
\begin{eqnarray}
R(z,x_{m})&=&\left( \sqrt{\frac{2}{1-z^{2}}}\right) \left( \sqrt{\frac{q^{2}(1+z)}{q^{2}(1+z)+k^{2}(1-z)}}\right) , \\
f(z,x_{m})&\sim & f_{0}(z,x_{m})=\frac{1}{\sqrt{\alpha z+\beta z^{2}}},
\end{eqnarray}
\begin{equation}
\alpha =1-\frac{2q^{2}}{q^{2}+x_{m}^{2}},
\end{equation}
\begin{equation}
\beta =\frac{q^{2}}{q^{2}+x_{m}^{2}}.
\end{equation}%
The radius $x_{m}$ of the photon sphere can be found from Eq.(2) as (it also follows from $\alpha =0$),
\begin{equation}
x_{m}=q\Rightarrow \beta _{m}=\beta |_{q=x_{m}}=\frac{1}{2},\; y_{m}=A(x_{m})=\frac{1}{2},
\end{equation}%
and the minimum impact parameter $u_{m}$ follows from Eq.(4) as
\begin{equation}
u_{m}=2q.
\end{equation}%
To compare the strong field lensing observables of this massless EMD wormhole with those of Schwarzschild black hole, we shall non-dimensionalize the parameters by the Schwarzschild radius $R_{\textmd{\scriptsize{s}}}$ as $q\rightarrow \frac{q}{R_{\textmd{\scriptsize{s}}}}$, $\Sigma \rightarrow \frac{\Sigma}{R_{\textmd{\scriptsize{s}}}}$, $R_{\textmd{\scriptsize{s}}}=2M_{\textmd{\scriptsize{s}}}$, where the subscript "s" stands for Schwarzschild. Then it follows from
Eqs.(18,19) that the exact coefficients, with ${\frac{2\beta _{m}}{y_{m}}=2,}
$ are
\begin{eqnarray}
\overline{a} &=&\frac{q}{\sqrt{k^{2}+q^{2}}}\Rightarrow \overline{a}=\left(
\frac{q}{R_{\textmd{\scriptsize{s}}}}\right) /\sqrt{\left( \frac{\Sigma }{R_{\textmd{\scriptsize{s}}}}%
\right) ^{2}+2\left( \frac{q}{R_{\textmd{\scriptsize{s}}}}\right) ^{2}} \\
\overline{b} &=&-\pi +b_{R}+\overline{a}\log {2}, \\
g(z,x_{m}) &=&-\frac{2q}{z\sqrt{1-z^{2}}}\left[ \sqrt{\frac{1-z^{2}}{%
k^{2}+q^{2}}}+\sqrt{\frac{1+z}{k^{2}\left( 1-z\right) +q^{2}\left(
1+z\right) }}\right] \\
b_{R} &=&\int_{0}^{1}g(z,x_{m})dz=-\frac{8q\log\left(q\right) }{\sqrt{%
k^{2}+q^{2}}} \\
&=&-\left[ 8\left( \frac{q}{R_{\textmd{\scriptsize{s}}}}\right) \ln \left( \frac{q}{R_{%
\textmd{\scriptsize{s}}}}\right) \right] /\sqrt{\left( \frac{\Sigma }{R_{\textmd{\scriptsize{s}}}}\right)
^{2}+2\left( \frac{q}{R_{\textmd{\scriptsize{s}}}}\right) ^{2}},
\end{eqnarray}%
For a correct comparison, the minimum impact parameter $u_{m}$ of rays in the Schwarzschild black hole and EMD\ wormhole spacetime should be the same, which implies%
\begin{eqnarray}
u_{m}^{\textmd{\scriptsize{Sch}}} &=&\left( \frac{3\sqrt{3}}{2}\right) R_{\textmd{\scriptsize{s}}}=\left( 3%
\sqrt{3}\right) M_{\textmd{\scriptsize{s}}}=u_{m}^{\textmd{\scriptsize{EMD}}}=2q \\
&\Rightarrow &\frac{q}{R_{\textmd{\scriptsize{s}}}}=\frac{3\sqrt{3}}{4}.
\end{eqnarray}
The last equation yields a formal identification of the Wheelerian mass $q$ with the BH\ mass $M_{\textmd{\scriptsize{s}}}$ as
\begin{equation}
q=\frac{\left( 3\sqrt{3}\right) M_{\textmd{\scriptsize{s}}}}{2}.
\end{equation}
The only variable in the Eqs.(38,39) now is the adimensionalized dilatonic charge $\frac{\Sigma}{R_{\textmd{\scriptsize{s}}}}$, and by varying it, we shall tabulate below the observables for massless EMD wormhole.

\section{Massless wormhole in EMS theory}
\label{Sec.4}
This massless wormhole is the zero Newtonian mass limit of the massive EMS wormhole \cite{17,18} sourced by an exotic scalar field $\phi$. The parent massive EMS wormhole metric is given by

\begin{eqnarray}
d\tau_{\textmd{\scriptsize{EMS}}}^{2} &=&Adt^{2}-Bdx^{2}-C(d\theta ^{2}+\sin ^{2}\theta d\varphi^{2})], \\
A(x) &=&\exp\left[-\pi\gamma +2\gamma\tan^{-1}\left( \frac{x}{m}\right)\right], \\
B(x) &=&A^{-1}(x),\quad C(x)=B(x)(x^{2}+m^{2}), \\
\phi(x) &=&\pm \kappa\left[\frac{\pi}{2}-2\tan^{-1}\left(\frac{x}{m}\right)\right],\quad 2\kappa^{2}=1+\gamma^{2},
\end{eqnarray}%
where $x\in (-\infty ,\infty )$, $m$ and $\gamma $ are arbitrary constants. This solution represents a static, horizonless, traversable, everywhere regular massive wormhole that has manifestly two asymptotically flat regions, one with Newtonian positive mass $M$ ($=m\gamma $) and the other with negative mass $-Me^{\pi \gamma }$, on either side of a regular throat at $x_{\textmd{\scriptsize{th}}}=M$ \cite{17}. The photon sphere has a radius $x_{m}=2M$. The metric (46) is indistinguishable from the Schwarzschild black hole as far as weak field predictions are concerned and that it can mimic the black hole with regard to the ring-down gravitational waves \cite{8}.

For the strong field deflection $\alpha $ in the EMS\ wormhole metric (46), we find that the integrand $g(z,x_{m})$ has a formidable expression that has to be integrated only numerically. Thus, we can in general write, using $x_{m}=2M=2m\gamma $,
\begin{equation}
b_{R}=\int_{0}^{1}g(z,x_{m})dz=\int_{0}^{1}g(z,M,\gamma )dz.
\end{equation}%
For the exact but complicated expression of $g(z,M,\gamma )$, see \cite{8}. It can also be verified for the metric (42) that
\begin{equation}
\log \left( \frac{2\beta _{m}}{y_{m}}\right) =\log \left[ \frac{\exp \left[
-4\gamma \tan ^{-1}(2\gamma )\right] \left[ \exp (\pi \gamma )-\exp \left\{
2\gamma \tan ^{-1}(2\gamma )\right\} \right] ^{2}\left[ 1+4\gamma ^{2}\right]
}{2\gamma ^{2}}\right] ,
\end{equation}%
The exact Schwarzschild black hole and the strong field coefficients can be deduced by choosing $\gamma =-i$ \ in Eqs.(18,19,50,51). Carrying out the numerical integration for $b_{R}$ for $\gamma =-i$, we obtain
\begin{eqnarray}
\overline{a} &=&1, \\
b_{R} &=&0.9496, \\
\overline{b} &=&-0.4002,
\end{eqnarray}%
which are just the known Schwarzschild values \cite{14}.

The massless EMS wormhole is defined by $M=m\gamma =0$. For this, one option is to choose $\gamma =0$ but $m\neq 0$ so that the metric reduces to
\begin{eqnarray}
d\tau ^{2} &=&dt^{2}-dx^{2}-(x^{2}+m^{2})\left( d\theta ^{2}+\sin ^{2}\theta
d\varphi ^{2}\right) , \\
\phi (x) &=&\pm \sqrt{\frac{1}{2}}\left[ \frac{\pi }{2}-2\tan ^{-1}\left(
\frac{x}{m}\right) \right] \simeq \pm \phi _{0}\pm \frac{\sqrt{2}m}{x}%
+O\left( \frac{m^{3}}{x^{3}}\right) ,
\end{eqnarray}%
which has been considered in \cite{7} for quasi-normal mode ringing. The other option $\gamma \neq 0$, $m=0$ yields trivial flat space and is of no physical consequence in our context.

The weak field deflection of light by this wormhole has been verified by three independent methods in \cite{26}. The meaning of $\pm m$ as Wheelerian masses, up to an unimportant constant factor of $\sqrt{2}$, is evident from Eq.(56). These masses can also be called scalar charges, since the total integrated energy of the ghost scalar field $\phi$ is $\pm m$. The mass $+m$ is responsible for inward bending of light on the positive side of the massless EMS wormhole. We find that%
\begin{equation}
\log \left( \frac{2\beta _{m}}{y_{m}}\right) =\log \left( \frac{\pi ^{2}}{2}%
\right),
\end{equation}%
and a much simplified expression%
\begin{equation}
g(z,0,0)=\frac{\pi z-\sqrt{2}\sqrt{1-\cos \left( \pi z\right) }}{z\sin
\left( \frac{\pi z}{2}\right) },
\end{equation}%
leading to
\begin{equation}
b_{R}=\int_{0}^{1}g(z,0,0)dz=\log (16)-2\log \pi .
\end{equation}%
Collecting the values from Eqs.(52), (57) and (59), we find
\begin{eqnarray}
\overline{b} &=&-\pi +b_{R}+\overline{a}\log \frac{2\beta _{m}}{y_{m}}=-\pi
+3\log (2), \\
\overline{a} &=&1.
\end{eqnarray}%
Further, since $x_{m}=0$ in the coordinate of the metric (55), it follows from Eq.(4) that%
\begin{equation}
u_{m}^{\textmd{\scriptsize{EMS}}}=m.
\end{equation}%
To compare the strong field lensing observables of the Ellis-Bronnikov wormhole with those of Schwarzschild black hole, we impose that the minimum impact parameters $u_{m}$ be the same for both so that

\begin{eqnarray}
u_{m}^{\textmd{\scriptsize{Sch}}} &=&\left(\frac{3\sqrt{3}}{2}\right) R_{\textmd{\scriptsize{s}}}=\left( 3%
\sqrt{3}\right) M_{\textmd{\scriptsize{s}}}=u_{m}^{\textmd{\scriptsize{EMS}}}=m \\
&\Rightarrow &\frac{m}{R_{\textmd{\scriptsize{s}}}}=\frac{3\sqrt{3}}{2},
\end{eqnarray}%
which allows us to connect the Wheelerian mass $m$ with the black hole mass $M_{\textmd{\scriptsize{s}}}$ as
\begin{equation}
m=\left( 3\sqrt{3}\right) M_{\textmd{\scriptsize{s}}}.
\end{equation}%
There is no free variable like dilaton in this case so that the coefficients are fixed. We shall estimate the observables in the ensuing Table.

\section{Numerical estimates}
\label{Sec.5}
For numerical estimates of the strong lensing signatures, we choose as Schwarzschild black hole the SgrA* residing in our galactic center\footnote{Strictly speaking, SgrA*\ black hole has spin and for comparison of observables, the appropriate wormhole examples should\ also be the spinning ones. However, choosing SgrA* is not mandatory for our analysis, any known Schwarzschild black hole would do. We point out that Sgr A* has nonetheless been modeled purely as a Schwarzschild black hole in the literature \cite{27}.} but add that any other black hole would be good enough for comparison with wormhole signatures provided they share the same $u_{m}$. If any other black hole is chosen, then only $u_{m}^{\textmd{\scriptsize{Sch}}}/R_{\textmd{\scriptsize{s}}}$ would numerically change as would the corresponding values of $q/R_{\textmd{\scriptsize{s}}}$ for EMD wormhole and $m/R_{\textmd{\scriptsize{s}}}$ for EMS wormhole. We adopt the latest observed data pertaining to SgrA* from [28]: Mass $M_{\textmd{\scriptsize{s}}}=4.2\times 10^{6}M_{\odot }$, $D_{\textmd{\scriptsize{OL}}}=7.6$ kpc, which imply $R_{\textmd{\scriptsize{s}}}=2M_{\textmd{\scriptsize{s}}}$, $u_{m}^{\textmd{\scriptsize{Sch}}}=\left( \frac{3\sqrt{3}}{2}\right) R_{\textmd{\scriptsize{s}}}$, $\theta _{\infty }=\left( \frac{u_{m}^{\textmd{\scriptsize{Sch}}}}{D_{\textmd{\scriptsize{OL}}}}\right) \times 206265\times 10^{6}$ $\mu $arcsec $=28.41$ $\mu $arcsec.

For Ellis-Bronnikov wormhole, $u_{m}^{\textmd{\scriptsize{Sch}}}/R_{\textmd{\scriptsize{s}}}\rightarrow m/R_{\textmd{\scriptsize{s}}}=\frac{3\sqrt{3}}{2}$ [see Eqs.(64,65)] so that the deflection angle is
\begin{equation}
\alpha (\Delta )=-\overline{a}\log \left( \frac{\Delta }{m/R_{\textmd{\scriptsize{s}}}}\right) +\overline{b},
\end{equation}%
where $\Delta =\left( u-u_{m}\right) /R_{\textmd{\scriptsize{s}}}$ represents the proximity of rays to the photon sphere (hence we call it the "proximity parameter") and ($\bar{a}$, $\overline{b}$) given by Eqs.(60,61).

For EMD wormhole, $u_{m}^{\textmd{\scriptsize{Sch}}}/R_{\textmd{\scriptsize{s}}}\rightarrow q/R_{\textmd{\scriptsize{s}}}=\frac{%
3\sqrt{3}}{4}$ [see Eq.(43,44)] so that
\begin{equation}
\alpha (\Delta ,\Sigma /R_{\textmd{\scriptsize{s}}})=-\overline{a}\log \left( \frac{\Delta
}{q/R_{\textmd{\scriptsize{s}}}}\right) +\overline{b}\left( q/R_{\textmd{\scriptsize{s}}},\Sigma /R_{\textmd{\scriptsize{s}}}\right) ,
\end{equation}

The table below shows comparison of strong lensing coefficients ($\bar{a}$, $%
\overline{b}$) and observables ($\theta _{\infty },s,r$) of wormhole lenses
composed of Wheelerian mass vis-\`{a}-vis SgrA* black hole lens of the same
minimum impact parameter $u_{m}$ as stipulated. To calculate the relevant
quantities, the algorithm is as follows: First we consider EMS wormholes,
both massive and massless. Putting the $x_{m}$ of wormholes in the master
Eqs.(18,19), we tabulate ($\bar{a}$, $\overline{b}$) in the 3rd and 4th rows
of the Table. Second, putting the adimensionalized Wheelerian mass $m/R_{\textmd{\scriptsize{s}}}=%
\frac{3\sqrt{3}}{2}$ in the expression for $\alpha $, we end up
with the remaining unknown quantity, the proximity parameter $\Delta $, in
Eq.(66). To fix it next, we note that the outermost image appears where $%
\alpha $ just equals $2\pi$,%
\begin{equation}
\alpha (\Delta )-2\pi =0.
\end{equation}

This equation is solved to find $\Delta $, which tells us how close the rays
should pass by the photon sphere in order to enable the wormholes to mimic
the deflection angle $2\pi $ caused by a black hole. However, the
observables ($\theta _{\infty },s,r$), determined solely by $\bar{a}$ and $%
\overline{b}$, do not show much difference from those of black hole except a
minor one in $s$. This means that the EMS wormholes, massive or massless,
can closely mimic a black hole in terms of strong field lensing observables.

The story with the EMD\ wormhole is quite different since there is now a
freely specifiable dilatonic charge $\Sigma /R_{\textmd{\scriptsize{s}}}$. Specifying
different values to it, we find that the equation, with the Wheelerian mass $q/R_{\textmd{\scriptsize{s}}}=\frac{3\sqrt{3}}{4}$, viz.,
\begin{equation}
\alpha (\Delta ,\Sigma /R_{\textmd{\scriptsize{s}}})-2\pi =0
\end{equation}%
corresponds to different sets $\Delta $ and of ($\theta _{\infty },s,r$) for
EMD wormholes, the difference signalling the presence of dilatonic charge in
the lens. These sets of values are markedly different from those of EMS
wormhole and black hole. The near zero values of $s$ for the EMD\ wormhole
show that the set of secondary asymptotic images will merge with the
outermost image producing a single image of the source, with the
characteristics that the flux ratio $r$ gradually increases as the parameter
$\Delta $ decreases or equivalently as the rays pass gradually closer to the
photon sphere.

\begin{table}[h]
\centering
\caption{ {\protect\footnotesize {The units in the table are: $\theta _{\infty }$ and $s$ in microarcseconds, $%
r$ in magnitudes. For the massless EMS and EMD wormholes, the
adimensionalized Wheelerian masses are $m/R_{%
\textmd{\scriptsize{s}}}=\frac{3\protect\sqrt{3}}{2}$ and $q/R_{\textmd{\scriptsize{s}}}=\frac{3\protect%
\sqrt{3}}{4}$ rrespectively so that $\protect\theta _{\infty }^{\textmd{\scriptsize{BH,EMS}}%
}=2\protect\theta _{\infty }^{\textmd{\scriptsize{EMD}}}$. The
values of $\Delta $ correspond to the same deflection angle $2\pi $ in all
cases, but the strong field lensing observables ($\theta _{\infty },s,r$)
are determined solely by $\bar{a}$ and $\overline{b}$ and are different for
different types of lenses.}}}
\label{Tab1}\tabcolsep=0.1cm
\begin{tabular}{|c|c|c|c|c|c|c|c|c|}
\hline
Lens & $u_{m}/R_{s}$ & - & $\Delta$ & $\bar{a}$ & $\bar{b}$ & $%
\theta_{\infty}$ ($\mu$as) & $s$ ($\mu$as) & $r$(mag) \\ \hline
SgrA* BH & $\frac{3\sqrt{3}}{2}$ & - & $3.25\times 10^{-3}$ & $1$ & $-0.40$
& $28.41$ & $0.035$ & $6.82$ \\ \hline
Massive & $\frac{3\sqrt{3}}{2}$ & - & $2.66\times 10^{-3}$ & $1$ & $-0.60$ &
$28.41$ & $0.029$ & $6.82$ \\
EMS WH &  &  &  &  &  &  &  &  \\ \hline
EMS WH & $\frac{3\sqrt{3}}{2}$ & - & $1.68\times 10^{-3}$ & $1$ & $-1.06$ & $%
28.41$ & $0.018$ & $6.82$ \\ \hline
EMD WH & $q/R_{\textmd{\scriptsize{s}}}$ & $\Sigma /R_{\textmd{\scriptsize{s}}}$ & - & - & - & - & - & -
\\ \hline
- & $\frac{3\sqrt{3}}{4}$ & $\pm 10$ & $2.95\times 10^{-33}$ & $0.12$ & $%
-3.32$ & $14.20$ & $3.2\times 10^{-32}$ & $4.23$ \\
- & - & $\pm 5$ & $5.26\times 10^{-18}$ & $0.24$ & $-3.48$ & $14.20$ & $%
5.7\times 10^{-17}$ & $3.53$ \\
- & - & $\pm 1$ & $8.22\times 10^{-8}$ & $0.62$ & $-4.01$ & $14.20$ & $%
9.0\times 10^{-7}$ & $2.51$ \\
- & - & $\pm 0.5$ & $3.21\times 10^{-7}$ & $0.68$ & $-4.09$ & $14.20$ & $%
3.5\times 10^{-6}$ & $2.41$ \\
- & - & $\pm 0.1$ & $5.11\times 10^{-7}$ & $0.70$ & $-4.13$ & $14.20$ & $%
5.6\times 10^{-6}$ & $2.37$ \\
- & - & $\pm 0.0$ & $5.21\times 10^{-7}$ & $0.71$ & $-4.13$ & $14.20$ & $%
5.7\times 10^{-6}$ & $2.37$ \\ \hline
\end{tabular}%
\end{table}

\section{Motion of timelike particles: stability of circular motion}
\label{Sec.5}
Although the main focus in this article has been the strong field lensing
observables caused by lightlike particles (rest mass zero), we shall for
completeness briefly touch upon the motion of timelike particles (rest mass
nonzero)\footnote{%
We thank an anonymous referee for drawing our attention to the motion of
timelike particles. This is a useful topic since the influence of secondary
gravitating sources on such particles has not yet been addressed in the
literature, to our knowledge.}. In particular, we shall examine here the
location of two asymptotically flat regions as required of a wormhole
geometry and the existence of stable circular orbits around the positive
mass side.

\textit{(a) EMD theory}

We start with the massless metric (29-31). To establish that it truly
represents a wormhole, we have to show that it has two asymptotically flat
regions on either side of the throat at $x_{\textmd{\scriptsize{th}}}=0.$ For convenience,
we put the metric (29-31) in the isotropic form ($t,R,\theta ,\varphi $) by
introducing the transformation
\begin{equation}
x=\frac{R^{2}-k^{2}}{2R},
\end{equation}%
which when inverted yields the "isotropic" throat radius $R_{\textmd{\scriptsize{th}}}=%
\frac{1}{2}x\left( 1\pm \sqrt{1+k^{2}/x^{2}}\right) \rightarrow k$ as $%
x\rightarrow x_{\textmd{\scriptsize{th}}}=0$. Discarding the negative sign, we find $\frac{R%
}{x}\rightarrow 1$ as $x\rightarrow \infty $, so at large distances $R$ and $%
x$ coincide. The metric (29-31) under the transformation (70) assumes an
isotropic form%
\begin{eqnarray}
d\tau ^{2} &=&A(R)dt^{2}-B(R)\left( dR^{2}+R^{2}d\theta ^{2}+\sin ^{2}\theta
d\varphi ^{2}\right)  \nonumber \\
&=&\left[ \frac{1}{1+\frac{4R^{2}q^{2}}{\left( R^{2}-k^{2}\right) ^{2}}}%
\right] dt^{2}  \nonumber \\
&&-\left[ \frac{k^{4}+4q^{2}R^{2}-2k^{2}R^{2}+R^{4}}{4R^{4}}\right] \left(
dR^{2}+R^{2}d\theta ^{2}+R^{2}\sin ^{2}\theta d\varphi ^{2}\right) .
\end{eqnarray}%
This metric is flat as $R\rightarrow \infty $ and is also invariant under
inversion: $R=\frac{k^{2}}{y}$, which implies the presence of another
asymptotically flat region at $y=0$. So, the metric form (71) represents a
twice asymptotically flat regular wormhole as the spacetime on either side
of the throat shows no curvature divergence at any point. Now redefine $R=2%
\overline{R}$ so that
\begin{equation}
d\tau ^{2}=A(\overline{R})dt^{2}-B(\overline{R})\left( d\overline{R}^{2}+%
\overline{R}^{2}d\theta ^{2}+\overline{R}^{2}\sin ^{2}\theta d\varphi
^{2}\right) .
\end{equation}%
Then the metric functions expand as%
\begin{equation}
A(\overline{R})=1-\frac{q^{2}}{\overline{R}^{2}}-\frac{q^{4}}{\overline{R}%
^{4}}\left( \frac{k^{2}}{2q^{2}}\right) +...
\end{equation}

\begin{equation}
B(\overline{R})=1+\frac{q^{2}}{\overline{R}^{2}}\left( 1-\frac{k^{2}}{2q^{2}}%
\right) +\frac{q^{4}}{\overline{R}^{4}}\left( \frac{k^{4}}{16q^{4}}\right)
+...
\end{equation}

Timelike test particles, with arbitrarily controllable parameters like
energy and angular momentum,\ are not relevant at least for lensing
observations. To understand what those particles would nevertheless see as
the "effective gravitating mass" of the massless wormhole, it would be
necessary to study the orbital precession of timelike test particles in both
the spacetimes. But if \textit{weak field} light deflection angle $\alpha $
is any indication, then note that the leading order deflection by the
massless EMD wormhole obtained in \cite{24} using the Gauss-Bonnet method:
\begin{equation}
\alpha (b)=\frac{\pi }{4b^{2}}\left( 3q^{2}-\Sigma ^{2}\right) ,
\end{equation}%
which reveals, following Schwarzschild formula, that the effective
gravitating mass is $M_{0}=\sqrt{3q^{2}-\Sigma ^{2}}$, and not merely $q$.
On the other hand, it can be verified that the PPN method of Keeton \&
Petters \cite{31} for the metric functions (73) and (74) using the gravitational
potential $\Phi =\frac{q}{\overline{R}}$ (and not $\Phi =\frac{M_{0}}{%
\overline{R}}$) also yield the same deflection (75)\footnote{%
Details omitted to save space but the calculation is straightforward.}
although the first order deflection term is absent anyway! So what is the
mass seen, $q$ or $M_{0}$? When dilaton is switched off, $\Sigma =0$, the
metric (29-31) reduces to the famous Einstein-Rosen bridge \cite{32} and in this
case, the mass is proportional to just $q$. While the intriguing question
raised above warrants a separate detailed investigation, it is expected that
both the parameters $q$ and $\Sigma $ would influence the motion of timelike
particles. A good example for demonstrating it is to investigate the
stability of circular orbits of timelike particles.

To this end, it is convenient to define $x^{2}+q^{2}=\rho ^{2}$ in metric
(29-31) so that the metric functions assume the "standard" coordinate
Morris-Thorne form \cite{33}
\begin{eqnarray}
d\tau ^{2} &=&A(\rho )dt^{2}-B(\rho )d\rho ^{2}-C(\rho )\left( d\theta
^{2}+\sin ^{2}\theta d\varphi ^{2}\right) , \\
A(\rho ) &=&1-\frac{q^{2}}{\rho ^{2}},B(\rho )=\frac{\rho ^{4}}{\left( \rho
^{2}-q^{2}\right) \left( \rho ^{2}+\Sigma ^{2}\right) },C(\rho )=\rho ^{2}.
\end{eqnarray}%
Henceforth, define affine parameter $\lambda $, use the four-velocity $%
U^{\mu}=\frac{dx^{\mu}}{d\lambda}$, and normalize the constants of motion
by $\epsilon $, viz., $E=U_{0}/\epsilon $, $L=U_{3}/\epsilon $. The geodesic
motion on the equatorial plane ($\theta =\pi /2$) has the equations
\begin{eqnarray}
A(\rho )\frac{dt}{d\lambda } &=&E,\rho ^{2}\frac{d\varphi }{d\lambda }=L, \\
g_{\mu \mu }U^{\mu }U^{\nu } &=&-\epsilon ^{2},
\end{eqnarray}%
where $\epsilon $ is the conserved rest mass of the test particle, $\epsilon
=1$ for timelike and $0$ for lightlike particles. The geodesic Eq.(79) can
be recast in the form%
\begin{eqnarray}
\frac{1}{2}\left( \frac{d\rho }{d\lambda }\right) ^{2}+V(\rho ) &=&\frac{%
E^{2}}{2}, \\
V(\rho ) &=&\frac{\epsilon ^{2}}{2}+\frac{L^{2}-q^{2}\epsilon ^{2}+\Sigma
^{2}\left( \epsilon ^{2}-E^{2}\right) }{2\rho ^{2}}  \nonumber \\
&&+\frac{L^{2}(\Sigma ^{2}-q^{2})-q^{2}\epsilon ^{2}\Sigma ^{2}}{2\rho ^{4}}-%
\frac{L^{2}q^{2}\Sigma ^{2}}{2\rho ^{6}}.
\end{eqnarray}

With the potential $V(\rho )$ at hand, the rest of the algorithm is quite
well known. At the constant critical radius $\rho =\rho _{c}$, one has $%
\left. \frac{d\rho }{d\lambda }\right\vert _{\rho =\rho _{c}}=0$ and $\left.
\frac{dV}{d\rho }\right\vert _{\rho =\rho _{c}}=0$, which provide $\rho
=\rho _{c}(q,\Sigma ,E,L)$ for $\epsilon =1$. If there is a stable circular
timelike orbit at $\rho =\rho _{c}$, then we should find $V^{\prime \prime
}(q,\Sigma ,E,L)\equiv \left. \frac{d^{2}V}{d\rho ^{2}}\right\vert _{\rho
=\rho _{c}}<0$. The expressions for $\rho _{c}$ and $V^{\prime \prime }$ are
rather large, hence omitted but what we find is that, for given spacetime
parameter values $\left( q_{0},\Sigma _{0}\right) $, and for \textit{some
choices }of orbital angular momentum $L_{0}$ (all in suitable units),
circular orbits with radii $\rho =\rho _{c}$ show $V^{\prime \prime }(\rho
_{c})<0$ for all values of energy $E$. Two typical Figs.1 and 2 are
exhibited for illustration.

\begin{figure}
  \centering
  \resizebox{0.75\textwidth}{!}{\includegraphics{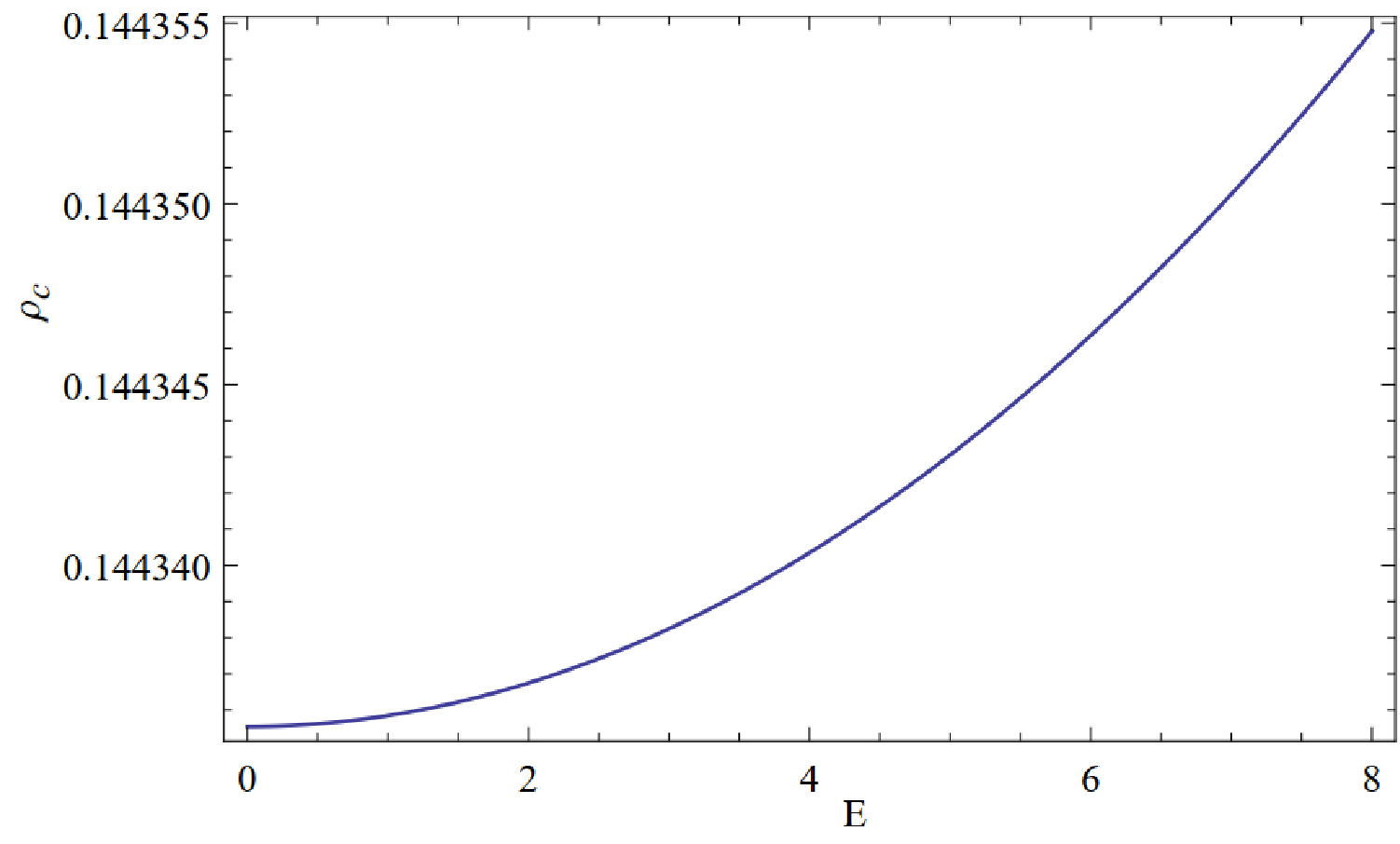}}
  \caption{Critical radius $\protect\rho =\protect\rho_{c}$ for the potential (80) for specific values $q=0.1$, $\Sigma =0.001$, $L=0.5$ (all in suitable units).}
\label{fig:1}
\end{figure}

\begin{figure}
  \centering
  \resizebox{0.75\textwidth}{!}{\includegraphics{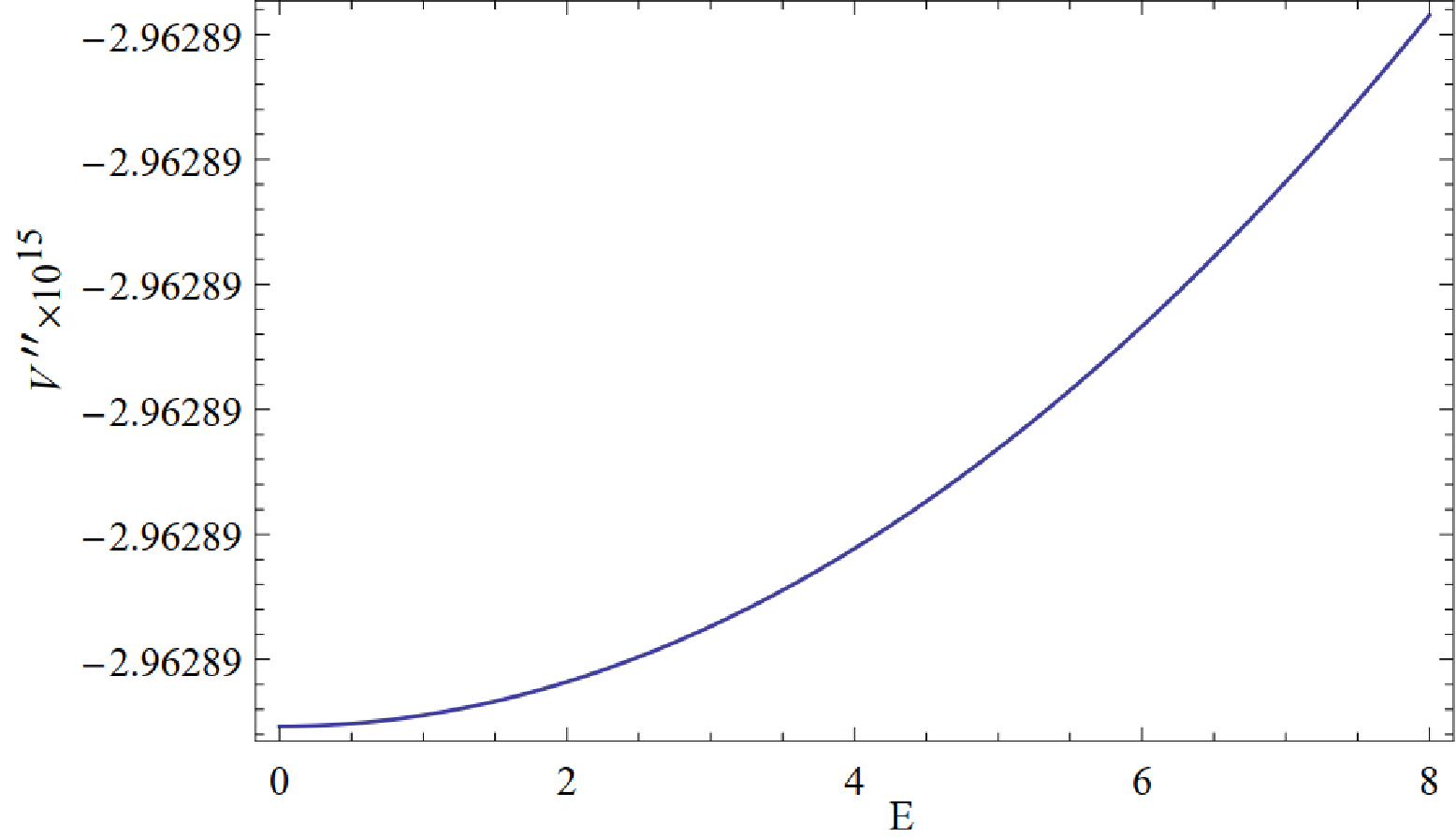}}
  \caption{Stability of circular radius $\protect\rho =\protect\rho _{c}$ for the potential (80) for specific values $q=0.1$, $\Sigma =0.001$, $L=0.5$ (all in suitable units), $V^{\prime\prime }(\protect\rho _{c})<0$.}
\label{fig:2}
\end{figure}

\textit{(b) EMS theory}

We start with the massless wormhole metric (55) that is covered on two sides
by the coordinate ranges $\infty <t<\infty $, $-\infty <x<\infty $, $0\leq
\theta \leq \pi $, $0\leq \varphi \leq 2\pi $. The metric is twice
asymptotically flat at $x\rightarrow \pm \infty $ since the curvatures
vanish there. For example, in the orthonormal frame (\symbol{94}) of an
observer, the Riemann curvature components are%
\begin{equation}
R_{\hat{\theta}\hat{\varphi}\hat{\theta}\hat{\varphi}} = -R_{\hat{x}\hat{\theta}\hat{x}\hat{\theta}} = -R_{\hat{x}\hat{\varphi}\hat{x}\hat{\varphi}} = \frac{m^{2}}{\left(x^{2}+m^{2}\right)^{2}},
\end{equation}%
which $\rightarrow 0$ as $x\rightarrow \pm \infty $. For the present
purpose, we cast the metric (55) in the Morris-Thorne "standard" form \cite{33}
using $x^{2}+m^{2}=r^{2}$, so that the metric on one side of the wormhole
manifold becomes ($-\infty <t<\infty $, $r>m$, $0\leq \theta \leq \pi $, $%
0\leq \varphi \leq 2\pi $):%
\begin{eqnarray}
d\tau ^{2} &=&A(r)dt^{2}-B(r)dr^{2}-C(r)\left( d\theta ^{2}+\sin ^{2}\theta
d\varphi ^{2}\right) , \\
A(r) &=&1,B(r)=\frac{1}{1-m^{2}/r^{2}},C(r)=r^{2}.
\end{eqnarray}%
The weak field light deflection angle $\alpha $, to leading order in the
massless EMS wormhole, is \cite{26}:%
\begin{equation}
\alpha (b)=\frac{\pi m^{2}}{4b^{2}},
\end{equation}%
which reveals, following Schwarzschild formula, that the "effective
gravitating mass" is just $m$. The orbital precession of timelike particles
should also involve only $m$, as there are no other extra solution
parameters.

Proceeding exactly as in (a), we find for timelike particles ( $\epsilon =1$%
) the potential
\begin{equation}
V(r)=\frac{1}{2}+\frac{L^{2}+m^{2}\left( E^{2}-1\right) }{2r^{2}}-\frac{%
L^{2}m^{2}}{2r^{4}},
\end{equation}%
which yields the critical circular radius $r_{c}$ and $V^{\prime \prime
}(r_{c})$:
\begin{eqnarray}
r_{c} &=&\frac{\sqrt{2}Lm}{\sqrt{L^{2}+E^{2}m^{2}-m^{2}}} \\
V^{\prime \prime }(r_{c}) &=&-\frac{\left[ L^{2}-m^{2}\left( 1-E^{2}\right) %
\right] ^{3}}{2L^{4}m^{4}}.
\end{eqnarray}%
$V^{\prime \prime }(r_{c})<0$ implies the stability condition%
\begin{equation}
L^{2}-m^{2}\left( 1-E^{2}\right) >0.
\end{equation}%
Thus stability is not decided by the mass $m$ alone, but also by the orbital
parameters $L,E$ of timelike particles that are to be so adjusted as to
satisfy the above inequality.

\section{Conclusion}
\label{Sec.6}
Throughout the paper, by the term massless, we mean that only the asymptotic
Newtonian mass is zero but the Wheelerian mass is not, so it can cause
observable lensing effects and stable circular orbits for timelike
particles. Since the nature of the mass of a black hole is not yet known, we
speculated that the same mass could as well be constituted by
scalar/dilaton/eletromagnetic fields. Although massless EMS wormhole has
been studied in the literature for its non-trivial lensing properties in
astrophysical applications, it is scarcely explained why, despite being
massless, it still bends light and exhibits\ observable lensing properties.

The present paper attempts to provide an explanation by invoking the
Wheelerian mass inherent in the massless wormholes and shows how its\textit{%
\ strong field lensing} properties compare with those of Schwarzschild black
hole, when both masses are quantitatively the same. Thus, for the massless
EMS wormhole, the asymptotic masses on either side, viz., $M$ ($=m\gamma )$
and $-Me^{\pi \gamma }$, could be zero for $\gamma =0$, but the Wheelerian
masses $\pm m$ need not be zero. The mass $+m$ is responsible for the inward
deflection of light and observable lensing effects as well as for the motion
of timelike particles. Likewise, the Wheelerian mass $q$ of the massless EMD
wormhole is provided by electric charge $P$ and magnetic charge $Q$ in a
combined form $q=\sqrt{2PQ}$ that is perceived in the observation of angular
radius of the shadow $\theta _{\infty }$, while the dilatonic charge $\Sigma
$ remains a free parameter. \textit{Our conclusion is that, while the
massless EMS wormhole can closely mimic a black hole (SgrA* chosen for
illustration) in terms of strong field lensing observables, the massless EMD
wormhole shows considerable deviation due to the presence of dilatonic
charge }$\Sigma $\textit{. Additionally, we showed that stable circular
orbits of timelike test particles exist in both massless spacetimes
depending on the choice of the orbital parameters. }

In a little more detail, we devised a proximity parameter $\Delta $, which
tells us how close the rays should pass by the photon sphere in order to
enable the wormholes to mimic the deflection angle $2\pi $ by a black hole
producing the outermost image, while the lensing observables ($\theta
_{\infty },s,r$) are determined exclusively by $\bar{a}$ and $\overline{b}$.
These observables\ are entered in the 4th row of the table, which show that
they are the same as those of black hole SgrA* except for a slight
difference\ in $s$.

The\ EMD\ wormhole strong field lensing observables differ quite
significantly from those of SgrA* black hole depending on the value of the
freely specifiable dilatonic charge $\Sigma $. Its presence in the lens
renders the values of image separation $s$ to be near zero as is evident
from the table. It indicates that the set of secondary asymptotic images
will merge with the outermost image producing a single image of the source,
with the characteristics that the flux ratio $r$ gradually increases as the
parameter $\Delta $ decreases or equivalently as the rays pass gradually
closer to the photon sphere. This feature seems to be a fundamental
characteristic of EMD\ wormhole that could constitute a potential test of
string theory.

The simplest strong field observable is the angular radius of the shadow $%
\theta _{\infty }$ \cite{29}, which alone can distinguish between massless EMS
and EMD\ wormholes since $\theta _{\infty }^{\textmd{\scriptsize{EMS wormhole}}}=2\theta
_{\infty }^{\textmd{\scriptsize{EMD wormhole}}}$. Remarkably, our table (2nd and 3rd rows)
shows that the measurement of $\theta _{\infty }$ cannot distinguish between
the black hole and a \textit{massive} EMS wormhole at the level of current
technology. A $\mu $as resolution is reachable in the next years by Very
Long Baseline Interferometry projects although there are lots of challenges
that would make the identification of the relativistic images very
difficult, as already enumerated in \cite{30}. Ongoing Event Horizon Telescope
project aims to achieve an accuracy of $\sim 15$ $\mu $as \cite{29}, obviously
still falling far short of the accuracy needed.

For completeness, we devoted Sec.6 to the study of timelike particles and
argued that stable circular orbits exist provided the orbital parameters are
suitably adjusted so as to satisfy the stability condition $V^{\prime \prime
}<0$. In addition, we had shown the EMD metric (71) is twice asymptotically
flat as required of a wormhole: One flatness is located at $R\rightarrow
\infty $ and the metric is also invariant under inversion: $R=\frac{k^{2}}{y}
$, which implies the presence of another asymptotically flat region at $y=0$%
. The asymptotic expansion of the metric is shown in Eqs.(73,74). Similar
properties of the EMS metric (55) are discussed in Sec.6(b). To answer what
the timelike test particles would see as the gravitating mass, it would be
more approriate to study the orbital precession of timelike test particles
in both the massless spacetimes in the potentials (81) and (86). Work is
underway.

\section*{Acknowledgement}

The reported study was funded by RFBR according to the research project No. 18-32-00377.

\end{document}